\documentclass[conference]{IEEEtran}
\newcommand{\ecrit}{\'{E}crit }

\usepackage{url}

\ifCLASSINFOpdf
\usepackage{todonotes}
  \graphicspath{{images/}}
  \DeclareGraphicsExtensions{.pdf,.jpeg,.png}
\else
\fi

\begin{document}

\title{Automated user documentation generation based on the
Eclipse application model}

\author{\IEEEauthorblockN{Marco Descher\IEEEauthorrefmark{1},
Thomas Feilhauer\IEEEauthorrefmark{1},
Lucia Amann\IEEEauthorrefmark{1}}
\IEEEauthorblockA{\IEEEauthorrefmark{1}Research Centre for Process- and
Product-Engineering\\
University of Applied Sciences Vorarlberg,
Vorarlberg, Austria\\ Email: [mde\textbar tf\textbar amlu]@fhv.at}}

\maketitle

\begin{abstract}
An application's user documentation, also referred to as the
user manual, is one of the core elements required in application distribution. 
While there exist many tools to aid an application's developer in creating and
maintaining documentation on and for the code itself, there are no tools 
that complement code development with user documentation for modern
graphical applications.

Approaches like literate programming are not applicable to this scenario, as not
a library, but a full application is to be documented to an end-user.
Documentation generation on applications up to now was only partially feasible
due to the gap between the code and its semantics. 

The new generation of Eclipse rich client platform developed applications is
based on an application model, closing a broad semantic gap between code and visible interface.
We use this application model to provide a semantic description for the
contained elements. Combined with the internal relationships of the application
model, these semantic descriptions are aggregated to well-structured user documentations that
comply to the ISO/IEC 26514.

This paper delivers a report on the \ecrit research project, where the
potentials and limitations of user documentation generation based on the
Eclipse application model were investigated.
\end{abstract}

\IEEEpeerreviewmaketitle

\section{Introduction}
\label{sect:introduction}

An application's documentation is one of the first ports of call for the user.
While there exist a myriad of tools and techniques for documentation generation 
targeted from the developer to the developer (developer documentation), 
documentation generation tools targeted at the application’s end-user (user
documentation) are barely available.

This is due to several facts and differences when it comes to generating user
and developer documentation:
\begin{enumerate}
  \item Developer documentation is targeted
  at single artifacts like classes and packages only, not considering any
  inter-relations\footnote{The Intent project \cite{INTENT} provides a
  framework for expressing such inter-relations.}.
    \item The user documentation has to describe the meaning of the program
  that emerges out of program execution (e.g., its visual interface and 
  interaction capabilities) while the developer documentation describes the
  meaning of the single implementation artifacts.
  \item The implementation artifacts (classes, packages etc.) are not
  aware of their structured meaning (e.g., class \texttt{ClassX} is
  the controller of model type \texttt{TypeA} or \texttt{TypeB}).
  \item The implementation classes are not aware of their semantic 
  (e.g., user interface, controller, \ldots) in the application.
\end{enumerate}

The new generation of Eclipse-based Rich Client Platform (RCP) \cite[p.
89]{AOSA2012} applications is developed on the basis of an application model
\cite[sect. 17]{vogellaEclipse4RCP}, which for the first time provides
information on the structure and inter-connections of the application's components, 
thus partially remedying facts 3 and 4. 

The application model is agnostic to the actual programming language of the
implementation classes, only referencing them. Therefore, there is no limit on
the implementation language used. This makes the results applicable to any other language or
application framework that adopts the Eclipse application model. Additionally, 
analysis on the application model can be performed without the application
actually running, by examining the respective model files only.

In \cite{Descher_2014} we introduced an approach to provide user documentation
on the basis of the Eclipse application model, outlined the technical and
scientific approach and identified four core issues to be analyzed in order to provide a meaningful 
proposition to the main research question: \textit{What are the potentials and
limitations of user documentation generation based on the Eclipse application
model?}

Generation of application software user documentation is currently
seen as a separate process with respect to the software development, potentially 
leading to differences between the user documentation and the actual
presentation and behavior of the application. As one of the main requirements to
documentation is its timeliness and synchronicity to the code, documentation
and development life-cycle have to somehow be dubbed making this an additional
topic considered in this paper.

This paper presents the \ecrit research project, covering both the technical
realization, the research approach, and the outcome on the stated
question.

\section{State of the Art}
\label{sect:sota}

Up to now, research on automatic documentation generation was mainly focused 
on developer documentation. Concepts like literate programming
\cite{Knuth84literateprogramming} and projects like I-Doc
\cite{Johnson94dynamicRegeneration} represent activities in this field.
Approaches that try to automatically generate end-user documentation,
like \cite{Paris-etal98-adcs} and \cite{6081814}, generate the
documentation by additional models or constructs that need to be
maintained independently from the application.

The Eclipse environment is one of the most popular development environments for
software. Generation 4 of Eclipse RCP development defines an application
model\footnote{The current version of the application model can be found on
\url{http://git.eclipse.org/c/platform/eclipse.platform.ui.git/tree/bundles/org.eclipse.e4.ui.model.workbench/model/UIElements.ecore}.},
as the base for the application's user interface. This new application model
 allows to derive the required information, by the afore mentioned approaches for documentation creation, 
remedying the requirement of additional, documentation specific, models.

The Eclipse application model features 37 different element types, which may be
arranged into the following categories:

\begin{itemize}
  \item \textit{Visual Adjustment} elements determine the composition of the
  user interface. This category contains elements such as \texttt{Part}, \texttt{Perspective},
  \texttt{Window}, \texttt{PartStack}, \ldots
  \item \textit{Action Initiation} elements are the visual representation of an
  action to be executed, such as a \texttt{MenuItem} or a
  \texttt{ToolItem}. They are embedded as visual elements, where e.g., a \texttt{ToolItem}
  is represented as a clickable icon (cf. figure \ref{fig:derivingUIcontext}
  showing the embedding of a \texttt{ToolItem} in a \texttt{Part} view menu).
  \item \textit{Action Execution} elements connect the abstract definition of
  actions within the application model to the actual implementation
  classes. Part of this are, among others, \texttt{Command}s and their
  respective \texttt{Handler}s.
  \item \textit{Dynamic Elements} provide dynamic instances of action initiation
  elements within the application.
  \item \textit{Extension Elements} allow for the extension of the application
  model. This enables one to interfere with any aspect of the application model
  that might require adaptation.
  \item \textit{Meta Elements} are used to express application model internal
  connections.
\end{itemize}

\begin{figure*}[ht]
\centering
\includegraphics[width=5in]{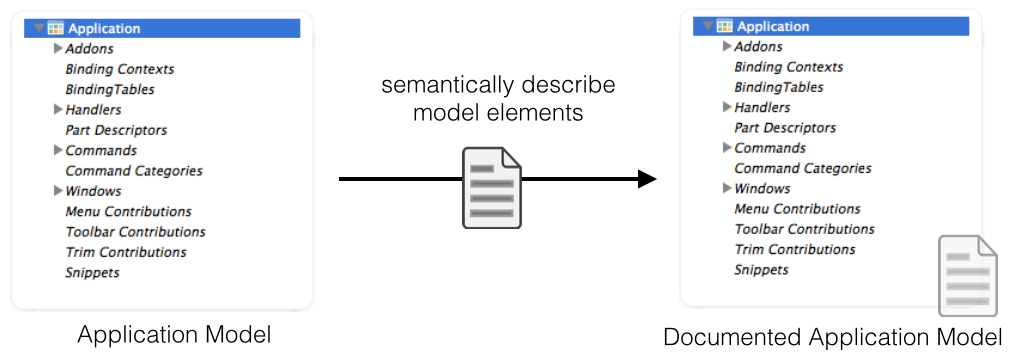}
\caption{Providing the semantic description of the model elements creates a
documented application model.}
\label{fig:documentedApplicationModel}
\end{figure*}

In the software development process, each application model element instance
is created due to a certain development requirement (such as a use case or a
user story). This development requirement describes a specific task to be
solved, which in turn contains the documentation on how to solve it (e.g., by
the formulated user story).
By tagging this information, the \textit{semantic description}, to the model element, a documented
model element is created. Assuming this has consistently been done to the entire
application model, the result will be a \textit{documented application model}
(cf. figure \ref{fig:documentedApplicationModel}).

The main standard specifying the structure and characteristic of a
software documentation is ISO/IEC 26514 \cite{ISO26514}. This specification sets
the objective for the generated user documentation, although the
approach should easily satisfy other documentation structures and/or means of
documentation.

\section{The Core Issues}
\label{sect:theCoreIssues}

In order to break down the main research question, stated in sect.
\ref{sect:introduction}, we identified four core issues \cite[sect.
4]{Descher_2014}. We briefly describe these core issues in this section and the
respective answer that was found in the course of the project.

\subsection{Mapping the application model to the documentation artifacts}
\label{sect:mappingAppModelDocArtifacts}
Given an ISO26514 \cite{ISO26514} conforming user documentation template,
we created based on the specification, the question was
\textit{whether the application model is currently expressive enough to 
allow for the mapping of application model elements to documentation elements}.

ISO26514 requires \cite[p. 46]{ISO26514} the following components to be
contained (in order) in the user manual

\begin{enumerate}
  {\color{gray}\item Identification data}
  {\color{gray}\item Table of contents}
  \item Introduction
  {\color{gray}\item Information for use of the documentation}
  \item Concept of operations
  \item Procedures
  \item Information on software commands
  \item Error messages and problem resolution
  \item Glossary
  {\color{gray}\item Navigational features}
\end{enumerate}

Components 1,2, 4 and 10 emerge out of the documentation
artifact to be created (e.g., \LaTeX \ documentation generated table of
contents), hence they are excluded from the core question - we refer to such
components as \textit{soft documentation elements}.

The remaining components require diverse information from the application model, so we have
to find separate answers. The general pre-condition for all components, however,
is that the application model has been documented (as described in sect.
\ref{sect:sota}) and that the application implements its tasks by representing
them within the application model. 

\begin{figure}[ht]
\centering
\includegraphics[width=3.5in]{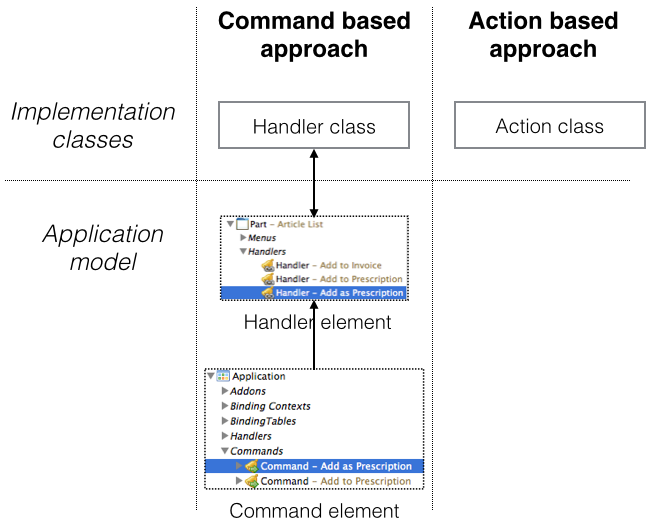}
\caption{To be visible within the application model, functionality has to be
implemented using the \texttt{Command} based approach.}
\label{fig:commandVsAction}
\end{figure}

That is, functionality within the application
may either be developed programmatically only, or by involving model elements.
One may either use a programmatic approach, i.e., use \texttt{Action} classes directly 
embedded into the user interface classes to solve a task, or define a
\texttt{Command} element within the application model, backed by its 
respective \texttt{Handler} class(es), represented in the user interface by a
contribution item. Both ways will lead to the functionality being available in the
application, yet only the \texttt{Command} approach leads to a model
visibility of the functionality's existence, as depicted in figure
\ref{fig:commandVsAction}.

\setcounter{subsubsection}{2}
\subsubsection{Introduction}
Has to contain soft documentation elements (e.g., document purpose, audience,
\ldots) and a brief overview of the software's purpose, functions, and operating
environment.

This component can not be satisfied by deriving information out of the model.
On the other hand, the introduction is a general description of the application
that can directly be derived from other documents of early software development 
phases, e.g., requirements or analysis documentations.

\setcounter{subsubsection}{4}
\subsubsection{Concept of operations}
Has to contain information on \textit{Software installation and uninstallation},
\textit{orientation to use the features of the graphical user interface} and \textit{navigation
through the software to access and to exit from functions}.

This component can be  generated out of the application model, except for the
software installation sub-component, as this is part of the deployment and not
the development procedure.

The sub-component on orientation documents basic user interface usage and hence
is derivable through the nature of the development framework. If one, for example, 
uses the Eclipse RCP framework for development, parts of the
framework documentation may be re-used in this component.

The sub-component on the navigation can be derived from the application model,
as it contains all the \texttt{Perspectives}, which in turn contain \texttt{Parts} describing the
single views visible to the user (cf. figure \ref{fig:derivingUIcontext}).

\subsubsection{Procedures}
A procedure is an ordered series of steps that specify how to perform a
task. This component describes the handling of complex tasks that require
multiple software commands, as described in the following section
(\ref{sect:infoOnSoftwareCommands}), to be executed in order.

There exists the programmatic concept of cheat sheets \cite[p.
165]{Reichert2009}. This feature, however, has not yet been ported to the
application model. Hence, this sub-component is currently not satisfiable out of
the application model.

\subsubsection{Information on software commands}
\label{sect:infoOnSoftwareCommands}
This component has to provide information on the commands, respective 
\texttt{Actions}, available. For each command it has to be documented what its parameters, 
pre-requirements and possible results (completion information or error
message) are.

As all commands are provided within the application model, this sub-component
can be satisfied. Given the semantic description of each command (cf. figure
\ref{fig:semanticDescriptionEditing}) a meaningful documentation can be
provided. By further analysis on the application model, it is also possible
to determine all initiators (that is the action initiation items as described
in sect. \ref{sect:sota}) of the respective command.

\subsubsection{Error messages and problem resolution}
Execution of commands may lead to error messages being presented to the user.
This component is to describe all possible presented error messages and
information on how to resolve the respective problem.

Error messages are not represented within the application model, hence it is
currently not possible to derive this section.

\subsubsection{Glossary}
The glossary has to describe all specialist vocabulary used in the specific
application. This information is not part of the application model, hence it is
not possible to be generated. Typically, a glossary is created anyway during the
analysis phase of the software development, and it can thus be used for the user 
documentation as well.

One may assume, however, that the application contains programmatic objects
modelling their real-world counterparts. If these objects are created using
a corresponding modelling technology, such as EMF \cite{EMF2009}, a big degree
of this component could be generated by parsing the underlying information of the
generated domain model.

\vspace{1em}
\textbf{Conclusion} While major parts of the documentation may already be
satisfied out of the given application model, it is not yet possible to satisfy
all requirements. Providing an extension to the application with respect to these
elements may however easily complete the generation of the user documentation.

The approach of a direct mapping between the application model and documentation
artifacts was refused, as the specification is too fuzzy in this
regard, and the requirements of such a model would be too concrete. We refer
to creating a document model for further processing instead, for details see
section \ref{sect:ecritPlugin}.

\subsection{Combining documentation artifacts and application model}
The development process, in the course of creating the application itself,
generates documentation artifacts (e.g., user stories) and meta-information on
the application context (e.g., domain knowledge).

These artifacts and meta-information have to be part of the application user
documentation.
To create a generable documentation, however, we have to find a formal connection between 
these artifacts and meta-information and the created application model elements (forming 
the application itself). In \cite[figure 4]{Descher_2014} we already identified
such a connection, namely the connection between \texttt{Action} (as defined in
\cite{ISO26514}) and \texttt{ActingEntity}.

The respective core issue here was \textit{to what extent is the connection of elements
in the application model and application documentation extendable}?

\begin{figure*}[ht]
\centering
\includegraphics[width=\textwidth]{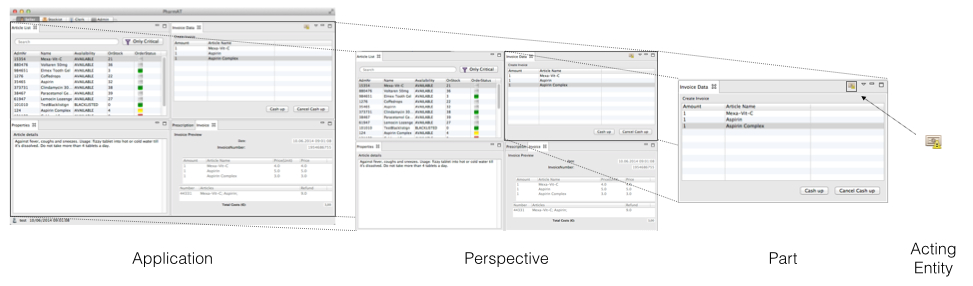}
\caption{Deriving user interface context information out of a single
(\texttt{ActingEntity},\texttt{Action}) connection}
\label{fig:derivingUIcontext}
\end{figure*}

During the course of the project we found out that the extension of the mapping
is not relevant, as it emerges out of the embedding of the already 
identified \texttt{ActingEntity} connection within the application. That is, cf.
figure \ref{fig:derivingUIcontext}, 
given a connection of (\texttt{ActingEntity},\texttt{Action}) the specific
action is embedded in the form of an initiatable item (button or menu item) at a
specific place in the user interface.

So here only the grouping of actions may be of interest, but
this is not relevant for the documentation. 

\vspace{1em}
\textbf{Conclusion} The element, through its placement
in the application method, bears sufficient contextual information in order to
determine more connection information.

\subsection{Combining development and documentation}
The application model can be modified by providing model fragments which get
merged at runtime providing a \textit{combined application model}. So the question here
was \textit{how to combine the Eclipse RCP development with the documentation
development, considering the dynamic structure of a delivered product}.

In Eclipse RCP we have a set of plugins and features,
that eventually get combined into a product forming the main distribution
artifact (i.e., the final application). So given a set of different plugins it is
possible to create different products with different features resulting in separate products for which a
respective user documentation is to be created.

Both the main application model and the fragments carry their own semantic
description, so the combination into one model generates a new documented
application model. The approach thus does not differ from an application where
we only have a single model, with the additional merging step only.

\begin{figure}[ht]
\centering
\includegraphics[width=3.3in]{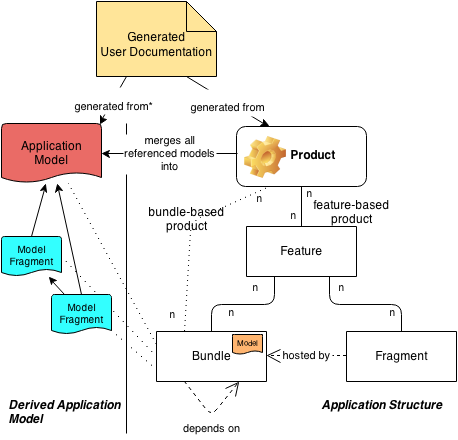}
\caption{User documentation can be created out of a single application model or
a product representing a combined set of functionality.}
\label{fig:dynamicDocumentationCombination}
\end{figure}

Figure \ref{fig:dynamicDocumentationCombination} shows that the original
approach, as presented in \cite[sect. 4.3]{Descher_2014} could basically be
adopted. A user documentation can either be created out of a single
main application model (this is not possible for a fragment) or a product. 

There exist several products containing only a single application model,
neglecting the possibility to use fragments. This may be due to an early design
phase, where no extension is required, or due to the nature of the project
itself. Allowing to create the documentation out of a single application model,
in addition to a combined application model generated out of a product, allows for
earlier integration of documentation creation in the development process.

As the semantic description is contained within the actual development
artifacts, they inherit its versioning. 

\vspace{1em}
\textbf{Conclusion} We effectively combined the
development with the documentation process, as the necessity to document an
element emerges at the time it is created. This led us to coining the term
Documentation Aware Development which will be further elaborated in section \ref{sect:DAD}.

\subsection{Determine the solution's applicability}
A programming framework aims to enable the creation of almost
any kind of software. So the question here is \textit{to what types of applications 
a sufficient result may be achieved in terms of a usable documentation}.

This question boils down to an objective and subjective part. Objective requirements 
on \textit{completeness} \cite[sect. 11.1]{ISO26514} and 
\textit{accuracy} \cite[sect. 11.2]{ISO26514} are easily verifiable, and can be
fulfilled using a generated documentation. The subjective perception of the
documentation by the user, however, can only be statistically determined, 
which was not feasible in the course of the project. The overall quality of the 
generated user documentation, however, depends to a very high degree on the
quality of the application model and its documentation.

\vspace{1em}
\textbf{Conclusion} Applicability measurement was changed to the objectively
determinable factors completeness and accuracy. To this end we conducted an
analysis on existing, Eclipse application model based, open source projects. 
For details please see section \ref{sect:analysisOnOpenSourceProjects}.

\section{Starting Point: Sample Application}

In order to treat the issues, as mentioned in sect. \ref{sect:theCoreIssues},
and to generate the toolkit for automated user manual generation, we needed a
sample application to derive the requirements from. To this end, the Pharmacy Austria
(PharmAT)\footnote{The PharmAT application and the referenced artifacts are
available at \url{https://github.com/ecrit/pharmacy_at}.} application was specified and
implemented.

The application is comprised of 14 user stories with 4 different user roles,
resembling a representative subset of the tasks in an Austrian pharmacy. During
the specification phase it became already clear which basic views and commands
would have to be presented to the respective application users (that is user
roles).

To harness this gathered information right away, a stub application model was
created, containing the set of views and commands, with their semantic
description, as required by the user stories. As for each user role a dedicated
perspective was designated, the views could already be assigned to their
respective perspective leading to a relatively advanced (with respect to the
development phase) and already documented application model.

With the template of an ISO26514 conform user documentation on the output site,
implementation of PharmAT was conducted. Information required by the template
was fed back to the input fields for semantic description in the application
model, finally leading to the current \ecrit toolkit as described in the
following section.

\section{The \ecrit toolkit}

The \ecrit toolkit \cite{ecritHomepage} consists of two separate software packages: An
extension of the Eclipse application model editor to allow adding the semantic
description and the actual project generating the documentation.

\subsection{Eclipse application model editor extension}

The Eclipse application model editor is a graphical editor to configure the
application model. It supports all the application model elements, and allows to
configure their respective properties. In order to support the semantic
description for every element contained in here, a horizontal extension
possibility had to be created \cite{e4HorizonalExtension}, allowing to provide
cross cut properties for the application model elements.

\begin{figure}[ht]
\centering
\includegraphics[width=3.5in]{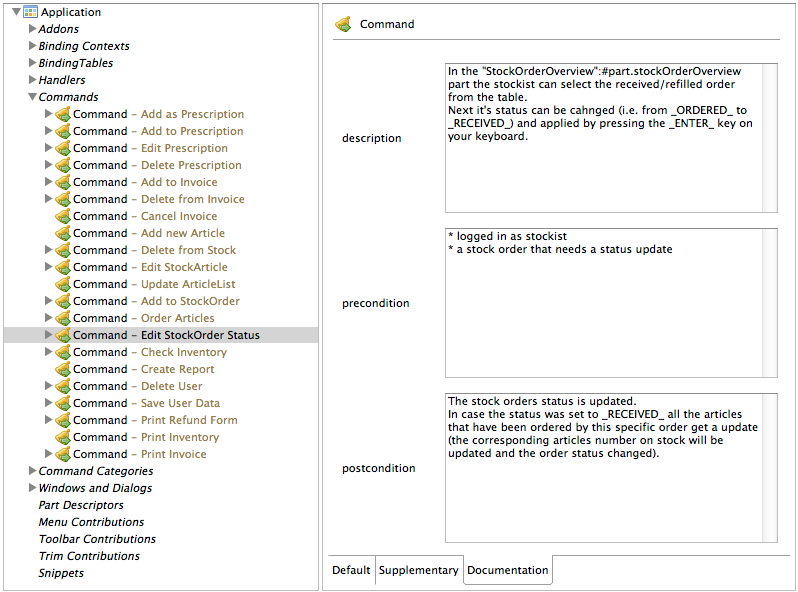}
\caption{Editing the semantic description of an application model command
element}
\label{fig:semanticDescriptionEditing}
\end{figure}

Depending on the model element to be edited, different properties have to be
presented in order to satisfy the documentation model requirements. Figure
\ref{fig:semanticDescriptionEditing} shows an example for the application model
element \texttt{Command}. Here the developer provides the \texttt{description} of the
command, the \texttt{precondition} for the commands execution and the \texttt{postcondition} after
executing the command. 

While \texttt{description} makes sense for almost any of the application, \texttt{pre-}
and \texttt{postcondition} do not make sense for non-action execution elements.
Considering other missing parts from the blank user documentation template also
leads to values that are valid for the entire application. These are for
example an \texttt{About}, where the user writes about the purpose of the entire
application, and some boolean values like \texttt{is multi-user} or
\texttt{requires login} that allow to consider respective documentation
sections.

The semantic description provided will be stored within the application model
respective application fragment itself, effectively binding it with the
development process.

\subsection{Eclipse \ecrit plugin}
\label{sect:ecritPlugin}

The \ecrit plugin tightly integrates into the development environment. By
executing it on an application model or a product element we
eventually receive a (combined) application model to generate the user documentation from.

\begin{figure}[ht]
\centering
\includegraphics[width=2.5in]{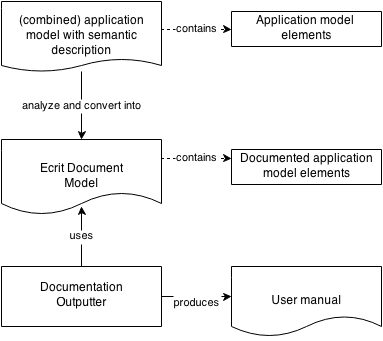}
\caption{The \ecrit user documentation creation process}
\label{fig:manualCreationProcess}
\end{figure}

Figure \ref{fig:manualCreationProcess} shows the subsequent process: The
(combined) application model is analyzed and transformed into the document
model. While in the application model all elements are contained as a tree, in
the document model they are ``enriched'' with information about their embedding.
That is, every element is informed about its children (applicable to visual
adjustment elements), where its executable from (applicable to action
exeuction elements, represented by action initiation elements), who is
referencing it, what groups it is part of, etc.

This is necessary to provide enough contextual information for the next step,
where the actual documentation artifacts are created. The outputter uses the
information contained in the document model to populate an ISO26514 conform
document template. Currently there exists corresponding HTML and \LaTeX\
outputters.

\begin{figure}[ht]
\centering
\includegraphics[width=3.5in]{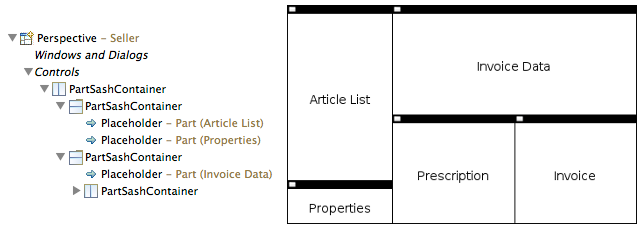}
\caption{Sample perspective in its model and depiction image representation.}
\label{fig:depictionImageWithModel}
\end{figure}

In addition to providing the required values for the outputter, the plugin also
creates depiction images for the perspectives available within the
application. A depiction image shows the structure of a perspective with the
arrangement of its parts in their relative position and the parts label for
identification. Fig. \ref{fig:depictionImageWithModel} shows such a depiction
image generated for a sample perspective. These images are to be embedded by the outputter in
the concept of operations section (cf. sect.
\ref{sect:mappingAppModelDocArtifacts}) of the generated user documentation.

The outputter itself may work on the provided documentation model in any way.
Currently we employ a simple text replacement system, where the provided
document template is filled with the required values. 

\begin{figure}[ht]
\centering
\includegraphics[width=3.5in]{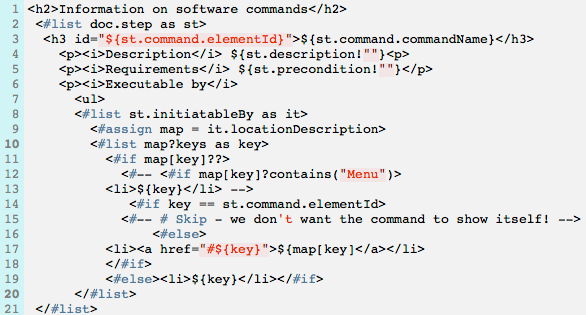}
\caption{Sample section of the HTML outputter template.}
\label{fig:templateSample}
\end{figure}

\begin{figure}[ht]
\centering
\includegraphics[width=3.2in]{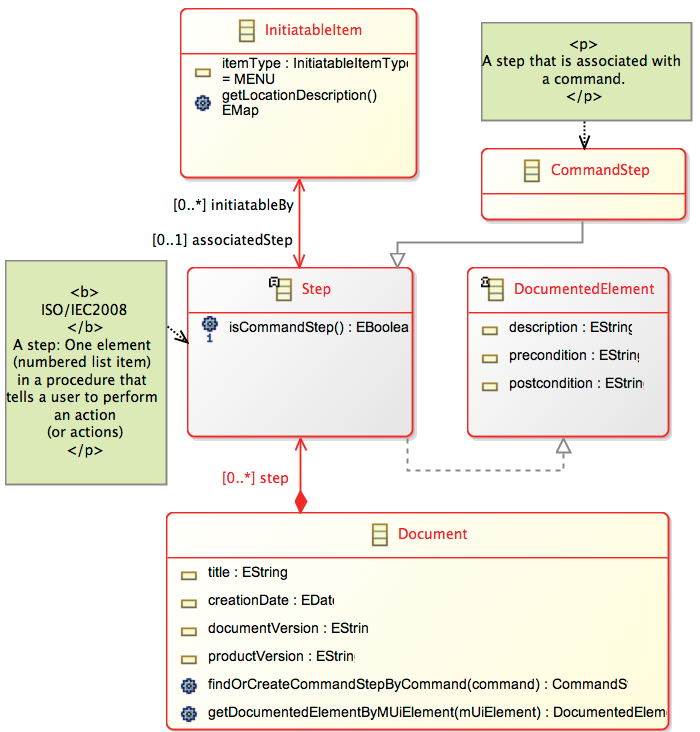}
\caption{Excerpt of the documentation model, showing the elements referenced
out of figure \ref{fig:templateSample}.}
\label{fig:docModelSample}
\end{figure}

Fig.
\ref{fig:templateSample} shows a section of the HTML outputter template, where
the commands documented in the application model get transformed into the
section information on software commands. The strings contained within the set
braces pre-pended by the dollar sign reference elements of the documentation
model as can be seen in figure \ref{fig:docModelSample}.

Given this basic approach, any required output could be generated, by
implementing the respective outputter. This is even encouraged by an extension
point provided by the \ecrit tooling on where to simply add an
outputter.  Details on the usage of the plugin can be found on the \ecrit
project page \cite{ecritHomepage}.

\section{Evaluation of the generated Documentation}
\label{sect:analysisOnOpenSourceProjects}

Due to the relative youth of the Eclipse 4 development platform, 
the amount of applications based on this system is limited.
To find a representative set of Eclipse 4 applications, we started with 
the three major open source software hosting sites according to 
\cite{OpenSourceHostingComparison}: google-code, sourceforge and github. 

As only github supports searches on the code repositories of all of its
projects, we limited the further analysis to this hosting platform. A
code search\footnote{Data collected August 8, 2014 accessible from
\url{https://github.com/ecrit/evaluation/raw/master/at.ecrit.github/rsc/result_08082014110324/toc.xls}}
revealed 395 repositories containing one or more application
model(s) or application fragments.

The majority of projects are of sample, educational or bug-fixing nature, with
no real value with respect to our analysis and/or very limited functionality.

To reveal applications with real-world content, we took the respective values of
our sample application as orientation to set the following selection criteria:

\begin{enumerate}
 \item In order for an application to be executable it has to have an
 application model, not a fragment only
 \item the model contains $\geq20$ command elements
 \item the model contains $\geq5$ part elements
\end{enumerate}

As it is also possible to implement the application's functionality without
employing application model elements, these boundary values also
define a certain confidence that the respective project favours the modeled
approach to the implementation only approach (cf. figure \ref{fig:commandVsAction}).

These criteria left us with 5 projects, to create a documentation from. The
\ecrit toolkit was successfully applied on all of the respective application
models.
As we do not have the respective semantic information on the product, however,
the semantic description on the application model is missing, which makes the generated 
documentation useless for real users of the application. 

This, however, shows the principal applicability of the \ecrit plugin to
suitable projects using the Eclipse 4 application model. Even though it is not a good 
practice for software development, \ecrit could also be used to add the semantic
description to the application's components ex post. We will discuss an
appropriate process integrating the provision of user documentation into an 
agile development process in section \ref{sect:DAD}.

Documentation quality is only partially an objective matter (as mentioned by
the ISO26514 requirements) and most of the perception of its quality
is dependent on the reader's prior knowledge. A clear assessment methodology
with respect to the output generated by the \ecrit tooling is yet owed.

A possible approach to go is to query the developers of the respective projects
to employ the \ecrit toolkit, providing the semantic description of their
application models, and then to subjectively assess on our side the time required to get familiar
with their applications. This approach would, however, be only valid by
employing a statistically significant number of projects and user candidates for
the respective application.

To this end we may conclude that while objective factors like completeness and
actuality can be measured by our approach, the real value of the generated
documentation for the end-user may only be empirically derived, by further
implementing the methodology and having a sufficient set of adopters.

\section{Documentation Aware Development}
\label{sect:DAD}

In this paper we have shown that large parts of the user documentation can be
automatically generated based on an appropriate documentation of the (Eclipse) 
application model. But this means, that the user documentation is dependent on 
the quality of the documentation of the application model which is tightly 
integrated into the development process of the corresponding software product.

ISO26515 \cite{ISO26515} describes the process of developing user documentation
in an agile environment, stating that \textit{Designing, developing, and
testing user documentation is greatly assisted by the presence of life-cycle documentation such as a documentation 
plan, system design document, system test plan, release records, and problem
reports}. To this end we want the user documentation to be implanted directly
into the core of the application, the application model.

Our approach allows to integrate the development of the user documentation into
the agile software development process more tightly, this means that parts of
the life-cycle documentation artifacts (e.g., user stories) can be reused for user
documentation purposes, and that user documentation is up-to-date with the
current application version. When \cite[sect. 5.3]{ISO26515} states that
\textit{The life cycle documentation items may not be formal or highly detailed documentation, but they are still useful in developing the user documentation.}
This means for our approach that the application model documentation has to be written 
with the user's view in mind. The central model elements relevant for the generation of 
the user documentation are the user interface elements (e.g., parts, toolbars, menus, commands) 
and these are exactly those relating to the user behavior.

\textit{Agile development methods frequently discourage the creation of detailed
engineering support documentation and detailed technical specifications. This means 
that technical writers often do not have source documentation from which to extrapolate 
feature details.} \cite[sect. 6.1]{ISO26515} With our approach we provide
exactly this documentation required by the technical writers, and we provide the documentation in a way that allows for automatic 
generation of the user documentation.

On the other hand, this means that the developers and the technical writers have
to work closer together, so that a high quality documentation can directly be inserted 
into the application model at the time the application model is created: \textit{Because the 
communication in agile development is face to face rather than through the use of 
detailed life cycle documentation, information developers should be a part of the agile 
development teams from the beginning of the sprint, as well as during the documentation 
of future changes for the project backlog.} \cite[sect. 6.3.1]{ISO26515} They
even suggest \textit{pair programming with the developer and technical writer}
\cite[sect. 7.2]{ISO26515} as a viable solution.

This meets a central claim of \cite[sect. 6.5]{ISO26515}: \textit{In a project
using agile development, ideally the user documentation is developed in parallel and to the same schedule as 
the software. This enables software to be regularly released to customers with 
sufficient documentation.}

It also shows the importance of the role of the technical writer during the whole 
software development process:

\textit{In agile development it is important that the development of the user documentation 
is part of the same processes as the software product life cycle, and performed 
in conjunction with development of the software. This enables the software and the 
user documentation to be tested, distributed, and maintained together. In agile development, 
the software cannot be considered complete without the production and validation of the 
associated user documentation.}

\textit{Agile development processes may impact the writers of user documentation in the
following ways:}

\begin{itemize}
  \item \textit{allows for involvement early in the development process}
  \item \textit{allows for influencing the software design, particularly of the user
  interface} \cite[sect. 7.1]{ISO26515}
\end{itemize}

Again, we see that the user interface and its documentation play a crucial role
for the user documentation. Therefore, the modelling of the user interface and
its documentation has to go hand in hand with the creation of the user documentation. 
This strong interrelation between user documentation and software design has been 
recognized by \cite[sect. 7.2]{ISO26515} when stating \textit{The user documentation may itself become the design 
specification for the product under development, where the design details the external 
design features applicable to the user, such as the user interface and steps to use it, 
and not the internal implementation of the software. Details of the external design and 
information about why and how the user should use the feature are required for development 
of code, testing, and production of the user documentation. This approach may have other 
effects, such as changing the order in which user documentation is produced or tested.}
We have gone one step further by automatically generating the user documentation based 
on the application model.

Altogether, this allows us to coin the term \textit{Documentation Aware Development}
which describes an agile software development process with a tight integration of 
documentation in all phases and the possibility to automatically generate sufficient 
and current user documentation.

\section{Conclusion}
In this paper we have presented an approach on how user documentation can automatically 
be generated from within software development projects. The implemented \ecrit
toolkit is a plugin for Eclipse 4. It allows to add semantic description to the components 
of the application model and enables the generation of accurate and consistent user documentation 
in different formats. 

This user documentation keeps pace with the progress of the development 
of a software product and follows the specification of ISO/IEC 26514. It therefore perfectly 
complements processes for agile software development.

While the application developer has to adopt a modelled approach to develop the
application, the advantage of having a substantial part of the user
documentation generated out of the generated application model, might weight out
the required learning. Developers starting with this new technology, might
already fully adopt the modelled approach and simply have to integrate the
process of providing the semantic description to the single elements into their
development activities.

A next step could be to find a way to represent error messages in the application model
and to extend \ecrit to also handle error message artifacts. Another possible
extension to \ecrit is the connection with Intent project. Work on this
is currently under way and will allow the generated documentation to become even
more dynamic as it will be possible to link the documentation with any kind of
development artifact.

We have planned to apply the current version of \ecrit to concrete
software development projects in small and medium enterprises (SME) and smoothly
adapt \ecrit to practical requirements of larger projects and to integrate users
with their real demands. Especially for SMEs not capable of sustaining a
specalized documentation unit, a standardization-conform pre-generated
documentation template would also reduce the time required to get familiar with user documentation
development requirements.

\section*{Acknowledgment}

This work is partially sponsored by The Austrian Research Promotion Agency (FFG)
under project number 840165.

\bibliographystyle{IEEEtran}
\bibliography{IEEEabrv,references}

\end{document}